\documentstyle[11pt,mrs2001,epsfig]{article}
\begin{document}
\title{The Three-Dimensional Power Spectrum Measured from 2MASS}

\author{B. ALLGOOD$^1$, G. BLUMENTHAL$^2$, J. R. PRIMACK$^1$}
\affil{$^1$Department of Physics, University of California, Santa Cruz
95064, USA} 
\affil{$^2$Department of Astronomy and Astrophysics, University
of California, Santa Cruz 95064, USA}

\begin{abstract}
We present the three-dimensional power spectrum of galaxy clustering
using measurements of the angular correlation function from the second
incremental release of the Two-Micron All Sky Survey (2MASS).  The 
angular positions of galaxies out to a limiting magnitude of $K_{s}
\le 14.0$ (508,054 galaxies) determine the correlation function.  
We consider a variety of estimators of the catalog's angular correlation 
function, and we find the Landy and Szalay (LS) estimator to be the 
most reliable choice.  Using the LS estimator, we find that the angular 
correlation function takes the form of a power law with a break.  
The angular correlation function is then used in combination with the 
selection function to determine the power spectrum by inverting Limber's 
equation.  We find good agreement with previous measurements
of the power spectrum for $0.04 \la k \la 0.1 h~{\rm Mpc}^{-1}$,
and evidence for a peak in the power spectrum around $k 
\sim 0.03 h~{\rm Mpc}^{-1}$.  For $k \ga 0.1 h~{\rm Mpc}^{-1}$ we 
find that the power spectrum is biased higher than other surveys, 
which can be understood as a color-based selection effect.
\end{abstract}

\section{Introduction}
The power spectrum $P(k)$ is one of the most important statistics
characterizing large scale surveys.  There are many ways of obtaining
the power spectrum.  Which method is chosen is mainly dependent on
the type of information collected in the survey and the geometry of
the survey coverage area.  2MASS is a photometric survey with some
spectroscopic follow-ups to help determine the luminosity function;
therefore we have mainly angular information and little radial
information.  We begin by determining the angular correlation
function, which is in itself an interesting statistic.  We then invert
Limber's equation using the method of Baugh \& Efstathiou 
\cite{bib:BE93} along with the selection function, and determine the 
three-dimensional power spectrum.  Finally, we use a jackknife
estimator to determine the bias and errors of the angular correlation
function and carry this through the inversion, folding in systematic
errors to ultimately estimate the errors in our determination of the
power spectrum.

\section{Two-Point Angular Correlation Function}
\begin{figure}
\plotone{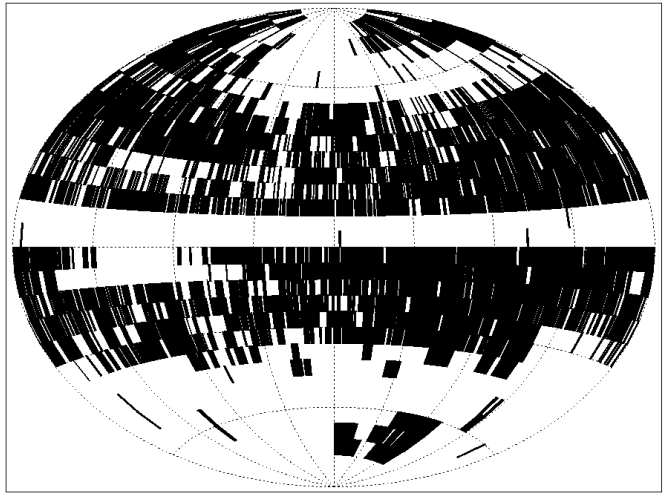}{3.5in}
\caption{An Aitoff plot in ra and dec of the mask 
determined from the 2MASS second incremental release catalog.  Note 
that the equator in this plot is not the galactic plane.}
\label{plot:mask}
\end{figure}

\begin{figure}
\plottwo{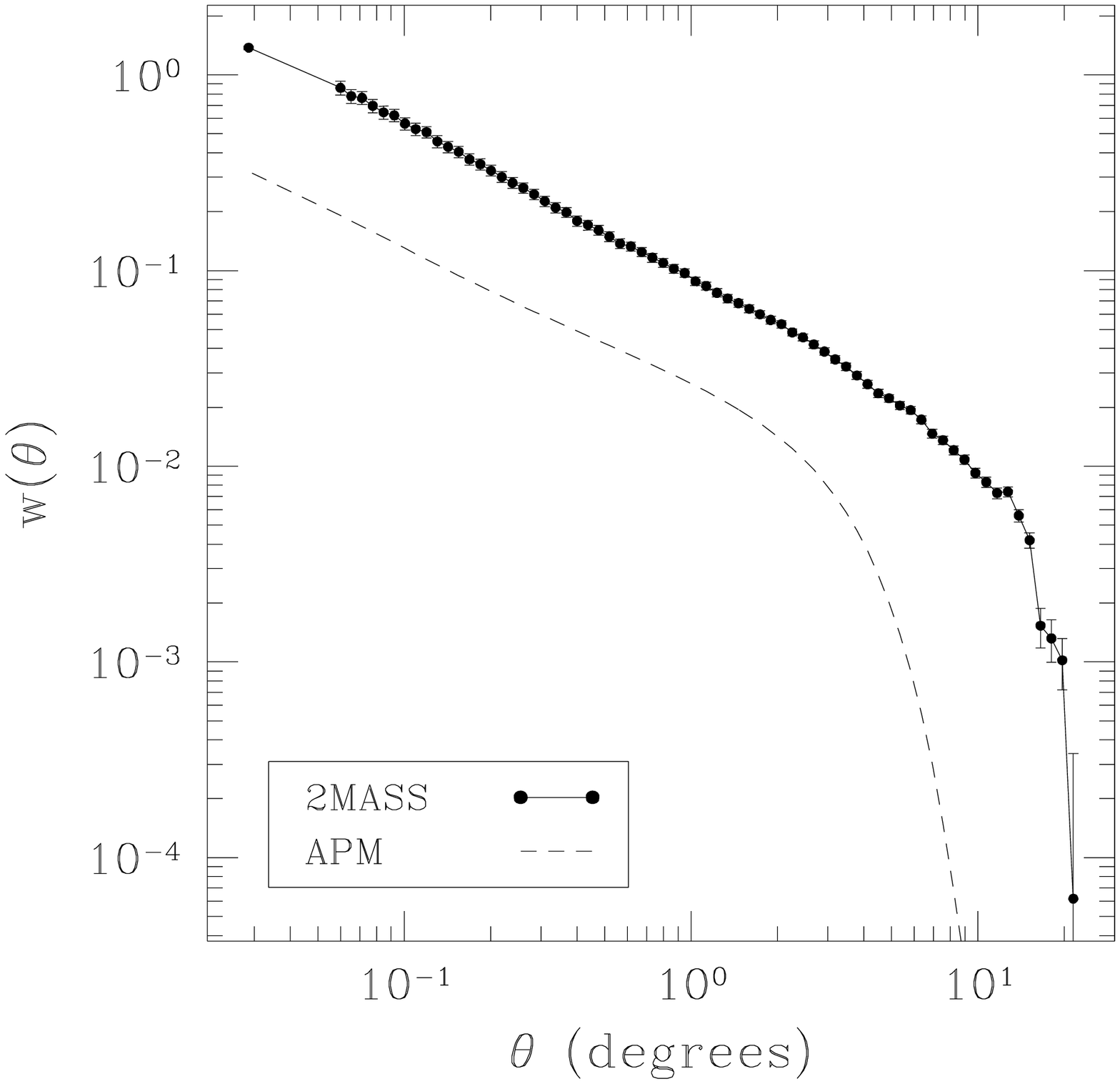}{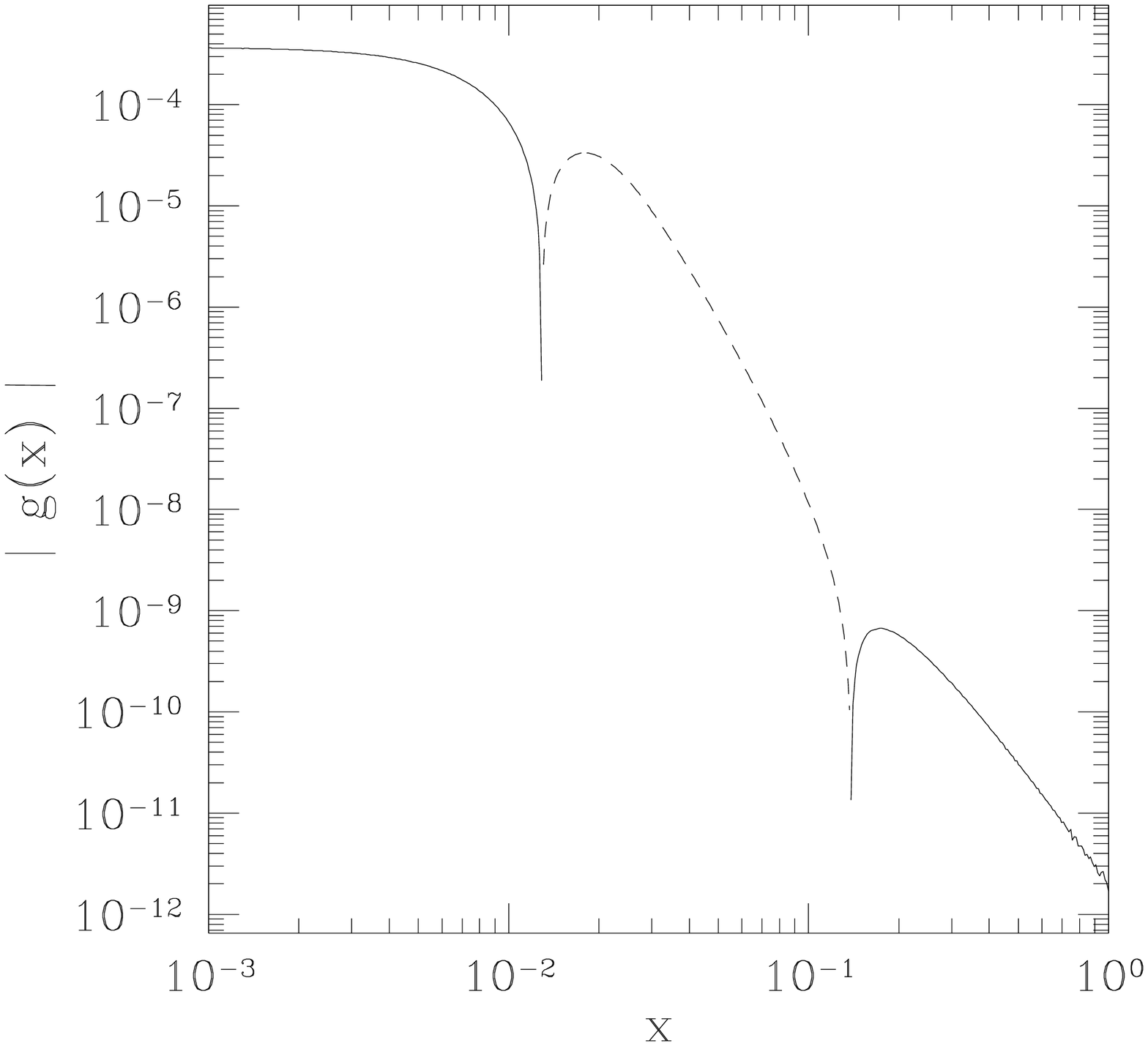}
\caption{The two-point angular correlation function and kernel from the
2MASS second incremental release.  Figure \ref{plot:ang-kern}a gives
the correlation function measured from the sample of galaxies with
limiting magnitude $K_{s} \le 14.0$.  The errors are calculated
using a resampling jackknife procedure and include Poisson errors
in the binning.  Figure \ref{plot:ang-kern}b is the absolute value of
the kernel calculated from the Schechter fit to the 2MASS K-Band
luminosity function.  In the dashed portion $g(x)$ is negative.}
\label{plot:ang-kern}
\end{figure}

The two-point angular correlation function $ w(\theta) $ is defined in
terms of the joint probability of finding two galaxies (out of a sample
of galaxies with angular density $ {\cal N} $) separated by an angle $\theta$ 
on the sky:
\begin{equation}
\delta P = {\cal N}^2\delta\Omega_1 \delta\Omega_2 [1+w(\theta_{12})]. 
\label{eq:2p_def}
\end{equation}
In practice one uses an estimator of $w(\theta)$ in order to deal with
the fact that the data are points and not a continuous function.
There are many standard estimators of the correlation function, all of
which rely on a Gaussian random field of points having the same
geometry as the data.  
In using these estimators one determines normalized bin counts of
data-data (DD), data-random (DR), and random-random (RR) pairs.  In
our analysis, we use logarithmic bins of angular separation in the
range of $0.05\deg \la \theta \la 11\deg$.  After testing the standard
estimators with a mock catalog having a known angular correlation we
find the Landy \& Szalay (LS) estimator \cite{bib:LS93}
\begin{equation}
w_{LS} = \frac{DD-2DR+RR}{RR}
\end{equation}
performs best at recovering the known correlation with our mask 
(Fig.~\ref{plot:mask}).
Other studies~\cite{bib:LSref2} of the standard estimators have also
shown the LS estimator to have the least variance.\\
Figure \ref{plot:mask} shows the mask for the galaxies in the 2MASS
second incremental release.  In order to avoid dust extinction from
the disk of the Milky Way we make a cut in galactic latitude.
We have calculated the angular correlation with cuts of $10\deg$,
$20\deg$ and $30\deg$ in galactic latitude and have found that these
different cuts produce very similar angular correlation functions.  
Therefore we use the angular correlation with the least conservative 
cut in order to improve statistics.  To further improve statistics we 
use many random samples with approximately two times the number 
of random points as data points to determine the angular correlation.  
The errors we determine for the angular correlation function include 
Poisson errors in the binning and variance in the angular correlation 
function from a jackknife resampling analysis.\\
We determine $w(\theta)$ (Fig.~\ref{plot:ang-kern}a) using a 
$O(N\log N)$ algorithm which involves an oct-tree decomposition.
The calculated $w(\theta)$ is of the expected form, with a power 
law on small angular
scales and a break between 1 and 10 degrees.  Using $\chi^2$ fitting
we find a power-law fit $\sim \theta^{-0.79}$, for $\theta <
1.7\deg$.  In comparison to APM~\cite{bib:BE93} and the latest angular
correlation from SDSS~\cite{bib:SDSSang}, the power laws are similar,
but the break is at much larger angular scales than the other two
surveys.  This is due to the fact that both SDSS and APM are much
deeper samples and therefore their angular correlation length is
expected to be smaller than that of 2MASS, assuming they have the same
three-dimensional correlation functions.

\section{Method}
Using the angular correlation function in combination with the
selection function $\phi(x)$, we need to invert Limber's equation
\begin{equation}
w(\omega)=\int_0^{\infty} P(k) g(k\omega) k dk,\;\;\;\omega = 2
\sin(\theta /2) \label{eq:limber}
\end{equation}
\begin{equation}
g(k\omega)=\frac{1}{2\pi N^2}\int_0^{\infty} x^4 \phi^2(x) J_o(k\omega x) dx \label{eq:kernel}
\end{equation}
to determine $P(k)$.  Since analytical solutions can only be found for
very few special cases, we perform the inversion numerically.
In particular, we use an iterative method of
inverting eq. \ref{eq:limber} which is based on Lucy's original
iterative method~\cite{bib:Lucy} modified to include non-positive 
definite kernels as done in ref.~\cite{bib:BE93}.  In
the algorithm one first makes an initial guess as to the shape of
the power spectrum and then convolves this with the kernel, thus
obtaining an angular correlation function.  One then calculates
multiplicative corrections to the power spectrum based on the
deviations from the measured angular correlation function.  This process
is then repeated until the power spectrum converges to a stable
solution and the computed angular correlation function matches the 
measured correlation function.

\subsection{Kernel}
The kernel (eq.~\ref{eq:kernel}) in Limber's equation
(eq.~\ref{eq:limber}) is determined from the selection function, which
is in turn dependent on the luminosity function and completeness.  We
use the best fit Schechter function parameters from Kochanek
et al.~\cite{bib:lum} for the luminosity function.  We ignore redshift effects on 
the colors as well as deviations from Euclidean geometry because of 
the shallowness of 2MASS (mean redshift of $z \sim 0.05$).  Figure
\ref{plot:ang-kern}b shows a plot of the kernel, where one should note
the two zero crossings across the range of $k\omega$ shown.  The first
zero crossing gives a natural scale for which the inversion process is
most sensitive $k\omega \sim 0.016h~{\rm Mpc}^{-1}$.  For a fixed value
of $\omega$ the integral is most sensitive to 
values of the kernel near this point.  Based on the 2MASS angular
correlation function (Fig. \ref{plot:ang-kern}a), we have sensitivity 
in the power spectrum down to wavenumbers of
order $k \sim 0.04 h~{\rm Mpc}^{-1}$ and up to $k \sim 10 h~{\rm
Mpc}^{-1}$.

\subsection{Tests of Inversion}
Tests on mock power spectra with the 2MASS kernel
show that the inversion has sensitivity down to $k = 0.02h~{\rm
Mpc}^{-1}$.  We find that the stability of the inversion becomes
very poor if we extend the range to include smaller wavenumbers.
Tests were also performed to determine how well the inversion
process is able to recover the angular correlation function.  
The recovered angular
correlation functions agree well with the mock measured angular
correlation functions, except in cases were the angular correlation
function contains a spike or sharp discontinuity.  For this reason 
we are unable to completely recover the sharp features in the measured 
angular correlation function (Fig. \ref{plot:ang-kern}a) 
in the vicinity of $\theta \sim 10\deg$.
By using an appropriate guess of the power spectrum in the first iteration 
for all of the test cases, the inversion method is able to converge to
a stable solution after only 10-30 iterations.  Allowing
the iteration to continue to 100-500 iterations yields little
change in the resulting $P(k)$ and no improvement in the recovered
angular correlation function.  Unlike the method presented in 
\cite{bib:BE93}, we found that there is no need for introducing a 
smoothing parameter.

\section{Power Spectrum}
\begin{figure}
\plotone{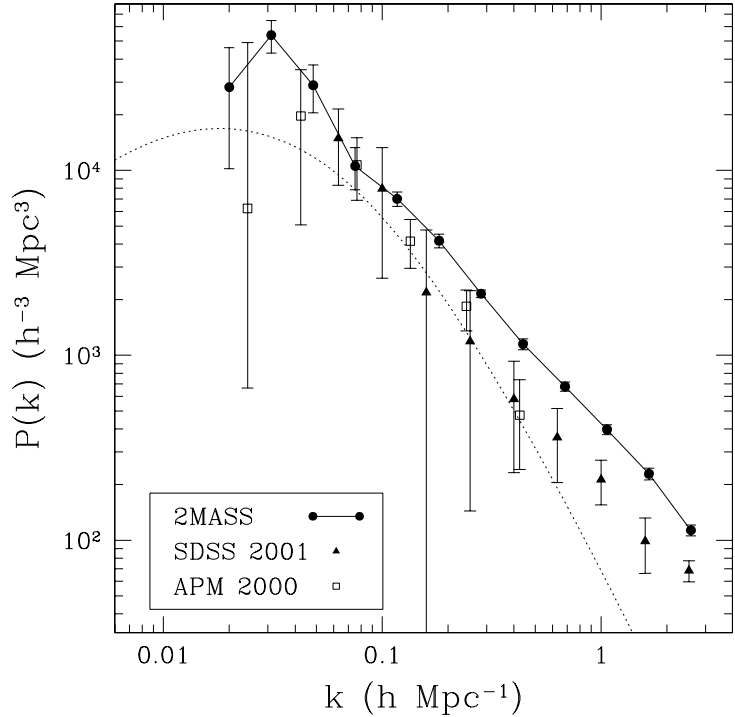}{4in}
\caption{The calculated power spectrum is shown above as filled
circles connect by a line.  Also plotted
are the recently released power spectrum from SDSS and the power
spectrum from the most recent analysis of the APM data.  The dashed
line is a no tilt $\Lambda$CDM linear power spectrum with
$\Omega_m=0.3$, $\Omega_b=0.04$ and $h=0.72$ using the
$\sigma_8 = 0.915$ from~\cite{bib:Szal_pow}.}
\label{plot:power}
\end{figure}

By applying the inversion method to the 2MASS angular correlation
function (Fig. \ref{plot:ang-kern}a) we obtain the power spectrum shown in
Fig. \ref{plot:power}.  Also shown in Fig. \ref{plot:power} are
the power spectra from the APM survey~\cite{bib:APMlatest} and SDSS
survey~\cite{bib:SDSSpow} with a no tilt $\Lambda$CDM linear power 
spectrum with $\Omega_m=0.3$, $\Omega_b=0.02$ and $h=0.72$.  The power 
spectrum we determine
from 2MASS shows the typical deviation from linear theory at large
values of $k$, but it is higher than that of APM and SDSS.  This
difference in the power between APM or SDSS and 2MASS on small
scales can be explained as a selection effect.  Both APM and SDSS
are samples of optically selected galaxies, while 2MASS is an infrared ($K_s$)
selected sample.  The selection of 2MASS will tend to pick out more
cluster galaxies than APM and SDSS and fewer field galaxies, therefore
showing more power on small scales.  There is fairly good agreement
between the surveys and the model in the wavenumber range of $0.05 \la
k \la 0.1 h~{\rm Mpc}^{-1}$, but even in this range the 2MASS power
spectrum has slightly higher amplitude.  This could mean that the power
spectrum of 2MASS has a slightly greater overall linear bias than the
other two surveys.  The relative bias between SDSS and 2MASS is $b
\sim 0.7-0.8$.  It is closer to $0.8$ if one tries to fit the model
spectrum or on the order of $0.7$ if the power spectra are forced to
match at $k = 0.252 h~{\rm Mpc}^{-1}$.  The most interesting feature
of the power spectrum is the very sharp peak at $k \sim 0.03 h~{\rm
Mpc}^{-1}$, which is also seen in the APM power spectrum.  The
error bars on the 2MASS power spectrum may be slightly optimistic, but
there is no doubt about the existence of this peak.  The width of this
peak is inconsistent with linear $\Lambda$CDM models and may be
showing us a new feature in the power, or possibly a contribution from 
baryons as claimed by 2dF~\cite{bib:2dF}.\\
The errors on the 
power spectrum take into account errors from the luminosity 
function, errors given for the angular correlation function, and 
estimates of the errors from the inversion technique.  The errors 
from the angular correlation function take into account the Poisson binning
errors and variance estimates from a jackknife resampling process.  There may be
errors introduced by the method used to determine the angular correlation 
which have not been accounted for.  One sigma errors from the luminosity 
function are also included.  To estimate the errors, we use different 
values for both the luminosity function and the angular correlation 
function within their respective errors.  We determine a statistically
large sample of power spectra using different values of these parameters.
From these power spectra, the extreme value for each point is 
multiplied by a correction factor, determined from the inversion, and is used
as the estimated error.  The errors for $k < 0.08 h~{\rm Mpc}^{-1}$ are 
small due to the fact that they depend on the angular correlations at 
small angular separation, which have corresponding small statistical errors.  They 
also seem to have less of a dependence on the errors in the luminosity 
function.  Due to the fact that the peak in the spectrum 
($k \sim 0.03 h~{\rm Mpc}^{-1}$) has such small error bars and that there 
are correlations in the errors due to the technique, the errors on the lowest 
$k$ point may be slightly underestimated.  Based on simulations done with 
mock power spectra the lowest $k$ point is otherwise believable.

\section{Conclusions}
In this paper we calculate the two-point angular correlation function
for the 2MASS second incremental release and obtain the
three-dimensional power spectrum using an iterative algorithm.  We find
that the 2MASS power spectrum recovered using this method is in good
agreement with power spectra from SDSS and APM, as well as model
predictions for $0.05 \la k \la 0.1 h~{\rm Mpc}^{-1}$.  For $k > 0.1
h~{\rm Mpc}^{-1}$ we find that the power spectrum of 2MASS deviates
sooner from the modeled linear spectrum and is higher than that of
SDSS and APM.  This is probably due to color selection.  We also find a
narrow peak in the power spectrum around $k \sim 0.03 h~{\rm
Mpc}^{-1}$, for which there is evidence in APM as well.  This could be the
result of baryonic oscillations in the power spectrum, or a new feature in 
the power spectrum which is yet not understood.  We plan to repeat this 
analysis as soon as the full 2MASS catalog becomes available.  This
will yield better statistics and will not be plagued by possible geometric
effects arising from the mask.

\acknowledgements{This publication makes use of data products from the
Two Micron All Sky Survey, which is a joint project of the University
of Massachusetts and the Infrared Processing and Analysis
Center/California Institute of Technology, funded by NASA and
NSF. We thank John Huchra for encouragement, discussions, and help.
This research was funded by grants at UCSC from NASA and NSF.}

\vfill
\end{document}